\documentstyle[12pt,colordvi,epsfig]{article}
\newcommand{\be}{\begin{eqnarray}}
\newcommand{\ee}{\end{eqnarray}}

\newcommand{\nd}{\noindent}
\newcommand{\jsi}{J/\psi}
\newcommand{\tilg}{\langle \tilde{\Gamma} \rangle}
\newcommand{\tilgx}{\langle \tilde{\Gamma} (x) \rangle}
\newcommand{\spt}{S(p_{{}_T})}
\newcommand{\stpt}{S_\perp(p_{{}_T})}
\newcommand{\sppt}{S_\parallel(p_{{}_T})}
\newcommand{\sopt}{S_0(p_{{}_T})}
\newcommand{\tlf}{t_{\mbox{life}}}
\newcommand{\pt}{p_{{}_T}}
\newcommand{\vsi}{\vec{v}_\psi}
\newcommand{\vel}{\vec{v}}
\newcommand{\ngx}{n_g(x)}
\newcommand{\ng}{n_g}

\newcommand{\lamg}{\lambda_{{}_g}}
\newcommand{\lamq}{\lambda_{{}_q}}
\newcommand{\lamqbar}{\lambda_{{}_{\bar q}}}
\newcommand{\vex}{v_{{}_e}}
\newcommand{\phii}{\phi^I_\psi}
\newcommand{\tsp}{t_{{}_{II}}-t_{{}_I}}

\begin{document}
\begin{center}
\medskip
\medskip
\medskip          
{\Large \bf $J/\psi$ Gluonic Dissociation Revisited : III.\\
Effects of Transverse Hydrodynamic Flow}
\vskip 0.2in
{\large B. K. Patra$^1$ and V. J. Menon$^2$}
\vskip 0.2in
{\normalsize{$^1$ Dept. of Physics, Indian Institute of Technology,
Roorkee 247 667, India\\
$^2$ Dept. of Physics, Banaras Hindu University, 
Varanasi 221 005, India}}
\end{center}
\vskip 0.3in

\begin{center}
{\bf Abstract}
\end{center}
In a recent paper [Eur. Phys. J {\bf C 44}, 567 (2005)] we developed a very 
general formulation
to take into account explicitly the effects of hydrodynamic flow profile
on the gluonic breakup of $J/\psi$'s produced in an equilibrating quark-gluon
plasma. Here we apply that formulation to the case when the medium is
undergoing cylindrically symmetric {\it transverse} expansion starting
from RHIC or LHC initial conditions. Our algebraic and numerical estimates
demonstrate that the transverse expansion causes enhancement of local
gluon number density $\ng$, affects the $\pt$-dependence of the average
dissociation rate $\tilg$ through a partial-wave interference mechanism, and 
makes the survival probability $\spt$ to change with $\pt$ very slowly.
Compared to the previous case of longitudinal expansion the new graph of
$\spt$ is pushed up at LHC, but develops a rich structure at RHIC, due to
a competition between the transverse catch-up time and plasma lifetime. 

\vskip 0.3in

\vskip 0.3in

\newpage
\section{\bf Introduction} 
It is a well-recognized fact that hydrodynamic expansion can significantly
influence the internal dynamics of, and signals coming from, the
parton plasma produced in relativistic heavy-ion collisions. The scenario
of $\jsi$ suppression due to gluonic bombardment~\cite{review}-\cite{gluon2} 
now becomes very
nontrivial because of two reasons: i) the flow causes inhomogeneities with
respect to the time-space location $x$ and ii) careful Lorentz 
transformations must be carried out among the rest frames of the fireball,
medium, and $\psi$ meson. In a recent paper~\cite{gluon2} this nontrivial
problem was formally solved by first assuming a {\it general} flow
velocity profile $\vec{v} (x)$ and thereafter deriving new statistical 
mechanical expressions for the gluon number density $n_g(x)$, average
dissociation rate $\tilgx$, and $\psi$ meson survival probability $\spt$
at transverse momentum $\pt$ (assuming the meson's velocity
$\vsi$ to be along the lateral $X$ direction in the fireball frame).

This general theory was also applied numerically in ref~\cite{gluon2} to
a plasma undergoing pure {\it longitudinal} expansion parallel to the
collision axis. In such case the kinematics is simple because $\vel \cdot
\vsi=0$ and also the cooling is known~\cite{dipali} to occur slowly. When
comparison was made with the no flow situation~\cite{gluon1} we found
that $\ngx$ was enhanced, a partial wave interference mechanism operated
in $\tilgx$, and the graph of $\spt$ was pushed down/up depending on
the LHC/RHIC initial conditions.

The aim of the present paper is to address the following important
question: ``{\it What will happen if the general theory of ref~\cite{gluon2} is 
applied to the cylindrically symmetric, pure transverse expansion
involving tougher kinematics (because $\vel \cdot \vsi \ne 0$) as well
as higher cooling rate}~\cite{dipali}? In Sec.2 below we derive the relevant
formulae for statistical observables ({\em viz}. $\ng$, $\tilg$, $\spt$, etc)
paying careful attention  to the $\psi$ meson trajectory and the so called
catch-up time. Next, Sec.3  presents our detailed numerical work along with
interpretations concerning $\tilg$ and $\spt$. Finally, our main
conclusions are summarized in Sec.4.
  
\section{\bf Statistical observables}
\setcounter{equation}{0}
\renewcommand{\theequation}{2.\arabic{equation}}
\subsection{\bf Hydrodynamic aspects}
We assume local thermal equilibrium and set-up a {\it cylindrical}
coordinate system in the fireball frame appropriate to central collision.
Let $\vec{x}=( r~\phi~z)$ be a typical spatial point, $x^\mu =(t,\vec{x})$
a time-space point, $\vel$ the fluid $3$ velocity, $\gamma={\left( 1 - v^2
\right)}^{-1/2}$ the Lorentz factor, $\tau$ the proper time,
$u^\mu=(\gamma, \gamma \vec{v})$ the $4$ velocity, $P$ the comoving pressure,
$\epsilon$ the comoving energy density, $T$ the temperature, and $T^{\mu \nu}
= \left( \epsilon +P \right) u^\mu u^\nu - P g^{\mu \nu}$ the energy-momentum
tensor. Then, the expansion of the system is described by the equation
for conservation of energy and momentum of an ideal fluid
\be
\partial_\mu T^{\mu \nu}=0 ,
\ee
in conjunction with the equation of state for a partially equilibrated
plasma of massless particles
\be
\epsilon=3P=\left[ a_2 \lamg + b_2 ( \lamq + \lamqbar) \right] T^4
\ee
where $a_2=8\pi^2/15$, $b_2=7\pi^2 N_f/40$, $N_f \approx 2.5$ is
the number of dynamical quark flavors, $\lamg$ is the gluon fugacity, and 
$\lamqbar$ ($\lamq$) is the (anti-) quark fugacity.
Of course, the gluons (or quarks) obey 
Bose-Einstein
(or Fermi-Dirac) statistics having fugacities $\lamg$ (or $\lamq$). Under
transverse expansion the fugacities and temperature evolve with the 
proper time according to the master rate equations~\cite{biro,munshi,jane}
\begin{eqnarray}
\frac{\gamma}{\lambda_g}\partial_t \lambda_g &+& \frac{\gamma v_r }{\lambda_g}
\partial_r \lambda_g +\frac{1}{T^3}\partial_t (\gamma T^3) + \frac{v_r}{T^3}
\partial_r (\gamma T^3)  \nonumber\\
&+& \gamma \partial_r v_r +\gamma \left( \frac{v_r}{r}+\frac{1}{t}\right)
\nonumber\\ &=&
R_3 ( 1- \lambda_g ) -2 R_2 \left( 1-\frac{\lambda_q \lambda_{\bar{q}}}
{\lambda_g^2}\right) \, ,
\\
\frac{\gamma}{\lambda_q}\partial_t \lambda_q &+& \frac{\gamma v_r }{\lambda_q}
\partial_r \lambda_q +\frac{1}{T^3}\partial_t (\gamma T^3) + \frac{v_r}{T^3}
\partial_r (\gamma T^3) \nonumber\\
&+& \gamma \partial_r v_r +\gamma \left( \frac{v_r}{r}+\frac{1}{t}\right)
\nonumber\\ &=&
R_2 \frac{a_1}{b_1} \left(
\frac{\lambda_g}{\lambda_q}-\frac{\lambda_{\bar{q}}}{\lambda_g}\right)\, ,
\end{eqnarray}
where the symbols are defined by
\be
&&R_2=0.5 n_g \langle v \sigma_{gg \longrightarrow q \bar{q}} \rangle, \quad
R_3=0.5 n_g \langle v \sigma_{gg \longrightarrow gg q} \rangle 
\ee
For our phenomenological purposes it will suffice to assume that, at a general
instant $t$ in the fireball frame, the plasma is contained in a uniformly
expanding cylinder of radius
\be
R=R_i+\left( t-t_i \right) v_e
\ee
where $R_i$ was the radius at the initial instant $t_i$ and the expansion
speed $v_e$ is a free parameter $(0 \le v_e < 1)$. In absence of azimuthal
rotations the transverse velocity profile of the medium can be parametrized
by a linear ansatz
\be
\vec{v}= v_e~\vec{r}/R, \qquad 0\le r \le R
\ee
Clearly, $|\vel|$ vanishes at the origin but becomes $\vex$ at the edge.
The (chemical) master equations (2.3 - 2.4) are designed to be solved 
numerically 
on a computer subject to the RHIC/LHC initial conditions stated in Table 1:

\begin{table}[h]
\caption{Colliding nuclei, collision energy, and initial parameters for the
QGP at RHIC(1), LHC(1)~\protect\cite{hijing}.}
\begin{tabular}{lccccccc}
\hline
& Nuclei & Energy $\sqrt{s}$ & $t_i$ & $T_i$ & $\lambda_{{}_{gi}}$ & 
$\lambda_{{}_{qi}}$ & $R_i$ \\
   &  & (GeV/nucleon) & (fm/c) & (GeV) & & &(fm)\\
\hline
RHIC(1)  & ${\mbox{Au}}^{197}$ & 200 & 0.7 & 0.55 & 0.05 & 0.008 & 6.98\\
  &  & & & & & &\\
LHC(1) & ${\mbox{Pb}}^{208}$ & 5000 & 0.5 & 0.82 & 0.124 & 0.02 & 7.01 \\
\hline
\end{tabular}
\end{table}
The lifetime or freeze-out time $t_{\mbox{life}}$ of the plasma is the
instant when the temperature at the edge falls to $T(t_{\mbox{life}})=0.2$
GeV, say.

\subsection{\bf Gluon number density}
For {\it arbitrary} flow profile $\vel$, momentum 
integration~\cite[eq.(11)]{gluon2}
over a Bose-Einstein distribution function yields the evolving gluon
number density
\be
n_g(x) = \frac{16}{\pi^2}~\gamma~T^3 \sum_{n=1}^\infty \frac{\lambda_g^n}{n^3}
\ee
Since this expression does not depend on the angles of $\vel$ it has the 
same structure both for the longitudinal and transverse cases. Also, flow
{\it enhances} the number density compared to the no-flow case~\cite{gluon1}; 
{\em e.g.} at fixed $\lamg$ the enhancement factor $\gamma$ becomes $2.3$ if
$|\vel |=0.9~c$.

\subsection{\bf Average $\psi$ dissociation rate}
In the fireball frame (keeping the flow profile still general) we consider a 
$\psi$ meson of mass $m_\psi$, four momentum $p^\mu_\psi$, three velocity
$\vsi=\vec{p}_\psi/p^0_\psi$, and Lorentz factor $\gamma_\psi=p^0_\psi/m_\psi$.
If $w^\mu$ is the plasma $4$ velocity measured in the {\it rest frame}
of $\psi$ then we can define the useful kinematic 
symbols~\cite[eq.(30)]{gluon2}
\be
&&F = \vec{v}\cdot \hat{v}_\psi, \qquad  Y = \gamma_\psi |\vec{v}_\psi| - 
\left(\gamma_\psi -1 \right)  F \nonumber\\
&& w^0 = \gamma \gamma_\psi \left( 1-F |\vec{v}_\psi| \right), \qquad 
\vec{w} = \gamma \left(\vec{v} - Y \hat{v}_\psi \right) \nonumber\\
&&\cos \theta_{\psi w} = \hat{w} \cdot \hat{v}_\psi
= \gamma \left(F-Y \right)/ |\vec{w}|
\ee
where the caps stand for unit vectors. Now, let $q^\mu$ be the gluon $4$ 
momentum seen in the $\psi$ meson rest frame, $\epsilon_\psi$ the
$c \bar c$ binding energy, $Q^0=q^0/\epsilon_\psi$ a dimensionless variable,
and $\sigma_{{}_{\rm{Rest}}}(Q^0) \propto {\left( Q^0 -1 \right)}^{3/2}/
{Q^0}^5$ the $g-\psi$ breakup
cross section according to QCD~\cite{BP}. Then, the mean dissociation rate 
due to hard thermal gluons~\cite[eq.(32)]{gluon2} is given by
\be
\tilgx &=& \frac{8 \epsilon_\psi^3 \gamma_\psi}{\pi^2}
\sum_{n=1}^\infty \lambda_g^n \int_1^\infty dQ^0 {Q^0}^2 \sigma_{\rm{Rest}} 
(Q^0) e^{-C_n Q^0} \nonumber\\
&\times& \left[ \frac{}{} I_0 (\rho_n) + I_1 (\rho_n) |\vec{v}_\psi| \cos 
\theta_{\psi w} \frac{}{} \right]
\ee
where we have used the abbreviations
\be
&& C_n = n \epsilon_\psi w^0/T, \qquad D_n = n \epsilon_\psi |\vec{w}|/T
\nonumber\\
&& \rho_n = D_n Q^0, \qquad I_0 (\rho_n) = \sinh (\rho_n)/\rho_n \nonumber\\
&&I_1(\rho_n) = \cosh (\rho_n) /\rho_n - \sinh (\rho_n)/\rho_n^2
\ee
Equation (2.10) demonstrates how $\tilgx$ depends on the hydrodynamic flow 
through $w^\mu$ as well as the angle $\theta_{\psi w}$. Retaining only the
$n=1$ term and picking-up the dominant peak contribution from $Q^0_p=10/7$
we arrive at the useful approximation
\be
&& {\langle \tilde{\Gamma}(x) \rangle} \propto \lambda_g \gamma_\psi H 
\nonumber\\
&& H \equiv e^{-C_1 Q^0_p} \left[ \frac{}{} I_0(D_1 Q^0_p)
+I_1 (D_1 Q^0_p) \mid \vec{v}_\psi \mid \cos \theta_{\psi w}
\frac{}{} \right]
\ee
in which a partial wave {\it interference} mechanism operates due to 
the anisotropic  $\cos \theta_{\psi \omega}$ factor. Numerical consequences
of (2.10) relevant to transverse flow will be discussed later in Sec.3.1.

\subsection{\bf $\jsi$ Survival probability}
In this section we shall consider pure {\it transverse} flow parametrized 
by (2.7)
and the $\psi$ meson moving in the {\it lateral} $X$ direction with velocity
$\vsi =(v_{{}_T}~0~0)$ appropriate to the mid-rapidity region in the fireball
frame. Suppressing the $z$ coordinate the production configuration of
$\psi$ meson is called $\left( t_I, \vec{r}^I_\psi \right) \equiv \left(
t_I, r^I_\psi, \phi^I_\psi \right)$ and the general trajectory after time
duration $\Delta$ is termed $\left( t, \vec{r}_\psi \right) \equiv \left(
t, r_\psi, \phi^I_\psi \right)$ such that
\be
&& t_I=t_i+\gamma_\psi \tau_{{}_F}, \qquad \Delta=t-t_I \nonumber\\
&& \vec{r}_\psi=\vec{r}^I_\psi+\vsi \Delta
\ee
where $\tau_F \approx 0.89~{\mbox{fm/c}}$ is the proper formation 
time~\cite{karsch}
of the $c \bar c$ bound state. This transverse trajectory will hit the edge
$R \equiv R_I + v_e~\Delta$ of the radially expanding cylinder ({\em cf.}(2.6))
at the {\it catch-up} instant $t^\ast$ after duration $\Delta^{^\ast}$ such 
that
\be
&& {\mid R_I + v_e \Delta^\ast \mid}^2 = {\mid \vec{r}^I_\psi + \vec{v}_\psi
\Delta^\ast \mid}^2 \nonumber\\
&& {\mbox{so}} \quad \alpha {\Delta^\ast}^2 + 2 \beta \Delta^\ast - \mu=0 
\nonumber\\
&& {\mbox{with}} \quad \alpha = v_\psi^2 -v_e^2, \qquad \mu = R_I^2- 
{r_\psi^I}^2 \nonumber\\
&& \beta  = r_\psi^I v_\psi \cos \phi_\psi^I -R_{{}_I} v_e 
\ee
If the quadratic in $\Delta^\ast$ has real roots we pick up that which is
positive and smaller; but if both roots are imaginary then catch-up cannot
occur. The time interval of physical interest becomes
\be
t_I \le t \le t_{II}, \qquad  t_{II} ={\mbox{min}} \left(t_I+\Delta^\ast,
~ t_{\mbox{life}}\right) 
\ee
This formula is quite different from that derived in the case of longitudinal
flow~\cite[eq.(48)]{gluon2}. As the time $t$ progresses the dissociation 
rate (2.10) must be evaluated on the $\psi$ meson trajectory itself, implying 
that we  have to set at a general instant
\be
&&\vec{r}=\vec{r}_\psi, \qquad \vec{v}= v_e~\vec{r}_\psi/R \nonumber\\
&&F \equiv \vec{v} \cdot \hat{v}_\psi = \left( \frac{v_e}{R}\right) 
\left( r_\psi^I \cos \phi_\psi^I + v_\psi \Delta \right)
\ee
in the kinematic relations (2.9). Clearly, the notation $\tilg$ of (2.10) 
becomes equivalent to
\be
\langle \tilde{\Gamma}[t] \rangle \equiv \langle \tilde{\Gamma} (t, 
p_{{}_T}, r_\psi^I, \phi^I_\psi ) \rangle
\ee
depending parametrically on the production configuration $r^I_\psi$, 
$\phi^I_\psi$. Then, by using the radioactive decay law without recombination
and averaging over the cross sectional area $A_I=\pi R_I^2$ (at the production
instant) we arrive at the desired survival probability
\be
&& S(p_T) = \int_{A_I} d^2 r_\psi^I (R_I^2 -{r_\psi^I}^2 ) e^{-W}/
\int_{A_I} d^2 r_\psi^I (R_I^2 -{r_\psi^I}^2 ) \nonumber\\
&& W = \int_{t_{{}_I}}^{t_{{}_{II}}} dt~~\tilde{\Gamma} [t], \qquad
d^2 r_\psi^I = d r_\psi^I~r_\psi^I~d \phi_\psi^I
\ee
Here no information is needed about the length $L_I$ of the cylindrical
plasma in contrast to the case of longitudinal flow~\cite[eq.(52)]{gluon2} 
where the averaging had to be done over the volume $V_I=\pi R_I^2 L_I$.

\section{\bf Numerical results} 
\setcounter{equation}{0}
\renewcommand{\theequation}{3.\arabic{equation}}
\subsection{\bf Curves of dissociation rate} 
The exact formula (2.10) of $\tilg$ is a very complicated function of $t$
as well as several kinematic parameters defined jointly by (2.9, 2.11, 2.16) 
but a feeling for its behaviour can be obtained in the extreme nonrelativistic
$\left( |\vel|/c \rightarrow 0 \right)$ and ultrarelativistic 
$\left( |\vel|/c \rightarrow 1 \right)$ 
limits. For simplicity, suppose at the {\it instant $t_I$} a special $\psi$
was formed almost at the edge $R_I$ of the cylinder with $\phi^I_\psi$
being the angle between the $\psi$ position vector and velocity
vector. Then the kinematic relations (2.16, 2.9) yield
\be
&& \vel= \vex \hat{r}^I_\psi= \pm \vex \hat{v}_\psi, \quad
F = \vex \cos \phi^I_\psi = \pm \vex \nonumber\\
&&\vec{w} = \gamma \left(\vec{v} - Y \hat{v}_\psi \right) 
= \gamma_e \left(\pm \vex -Y \right) \hat{v}_\psi
\ee
where the $+$, $-$ signs correspond to $\cos \phi^I_\psi = \pm 1$,
{\em i.e.}, to $\phi^I_\psi=0, \pi$, respectively. Thus we have the
parallel or anti-parallel property
\be
&& \vec{w} \parallel \hat{v}_\psi, \quad \cos \phi_{\psi w}=+1~{\mbox{if}}
~\phi^I_\psi=0~\&~Y < \vex \nonumber\\
&& \vec{w} \parallel - \hat{v}_\psi, \quad \cos \phi_{\psi w}=-1~{\mbox{if}}
~\phi^\pm_\psi=\pi~{\mbox{or}}~ Y > \vex 
\ee
Figures 1 - 4 depict the corresponding exact curves of $\tilg$ computed from 
(2.10) based on the LHC initial conditions of Table 1. We now proceed
\begin{figure}[htb]
\mbox{
\hskip1cm\epsfig{file=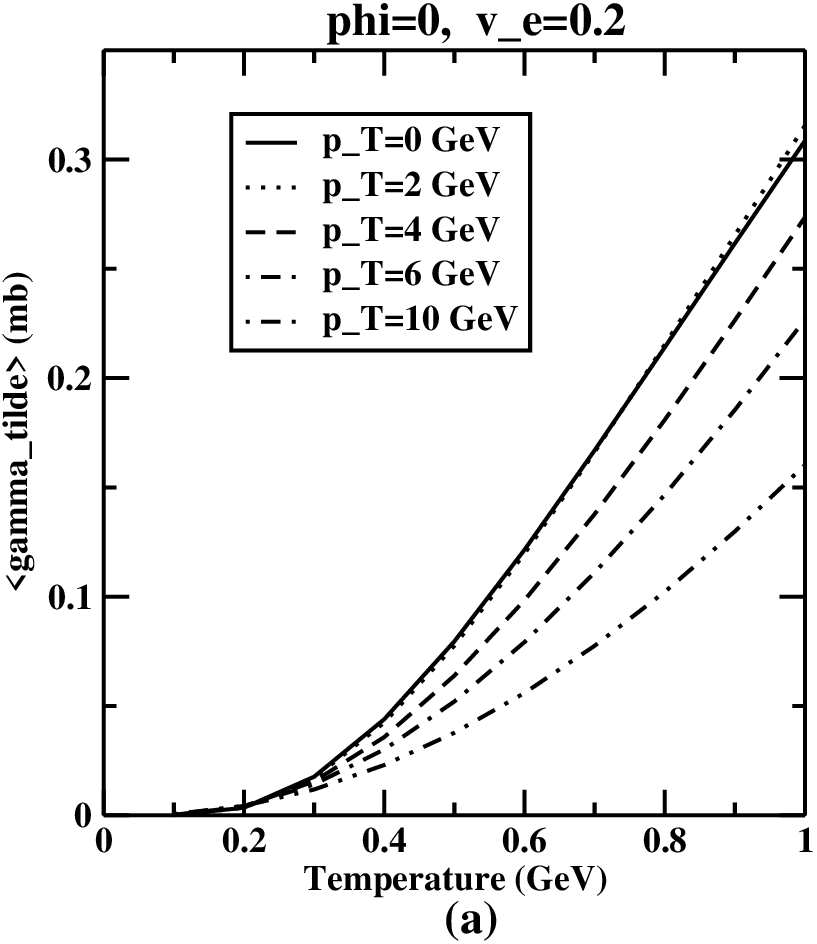,width=7cm}
\hskip1cm
\epsfig{file=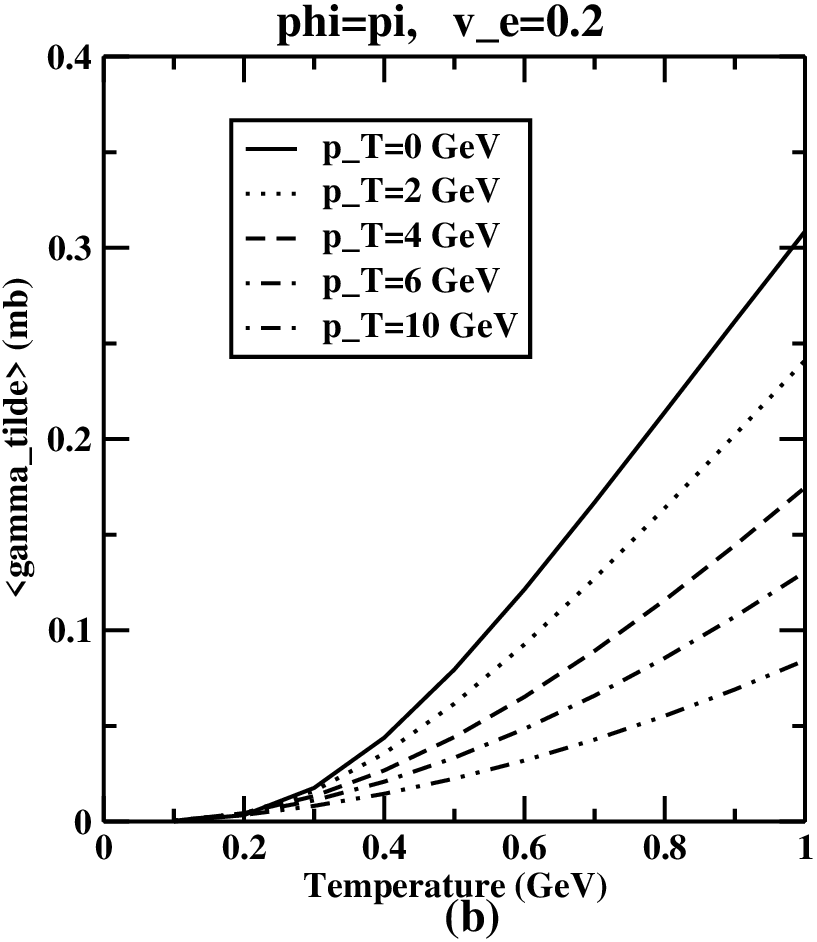,width=7cm}}
\vskip0.5cm
\caption{The variation of the modified rate $\tilg$ as a function of 
temperature at different transverse momenta for the transverse flow
velocity $v=0.2~c$ for (a) $\phii=0$ and (b) $\phii=\pi$, respectively.}
\end{figure}
to interpret these graphs using the approximate estimate (2.12).

\nd {\bf Interpretation}: i) At fixed $(\pt, \phi^I_\psi, \vex)$ the 
steady {\it increase} of $\tilg$ with $T$ in Figs.1 - 2 is caused 
by the growing $\exp\{ - (C_1 \mp D_1 ) Q^0_p \}$ factors occurring
in the estimate (2.12). 
\begin{figure}[htb]
\mbox{
\hskip1cm\epsfig{file=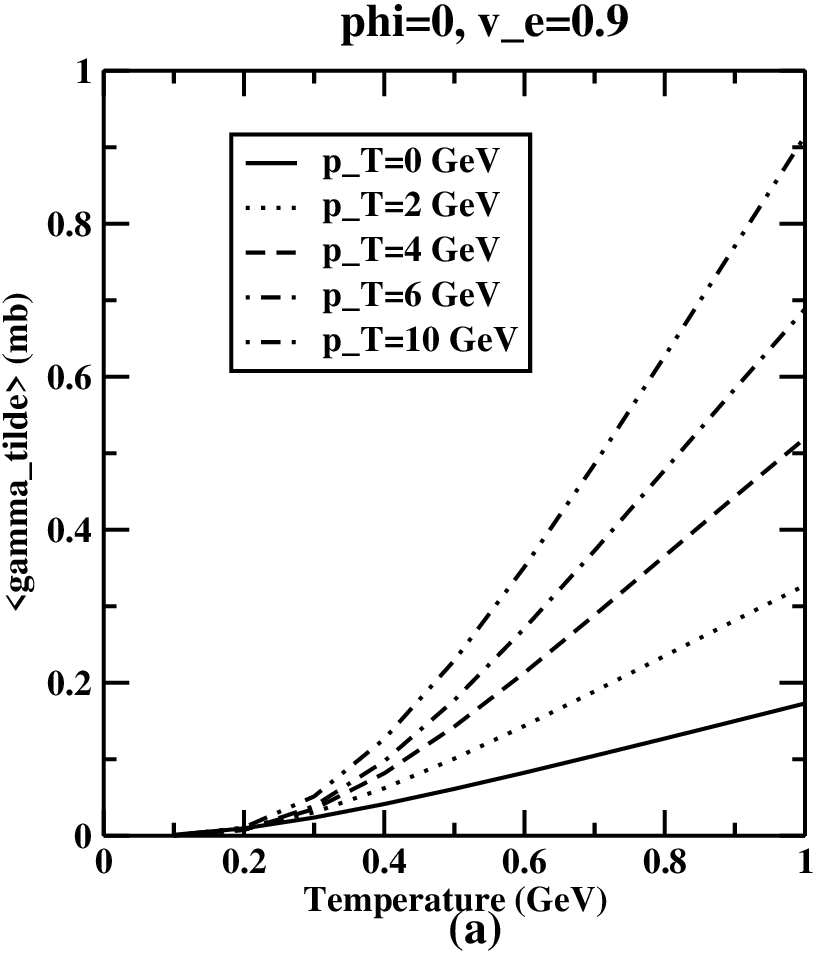,width=7cm}
\hskip1cm
\epsfig{file=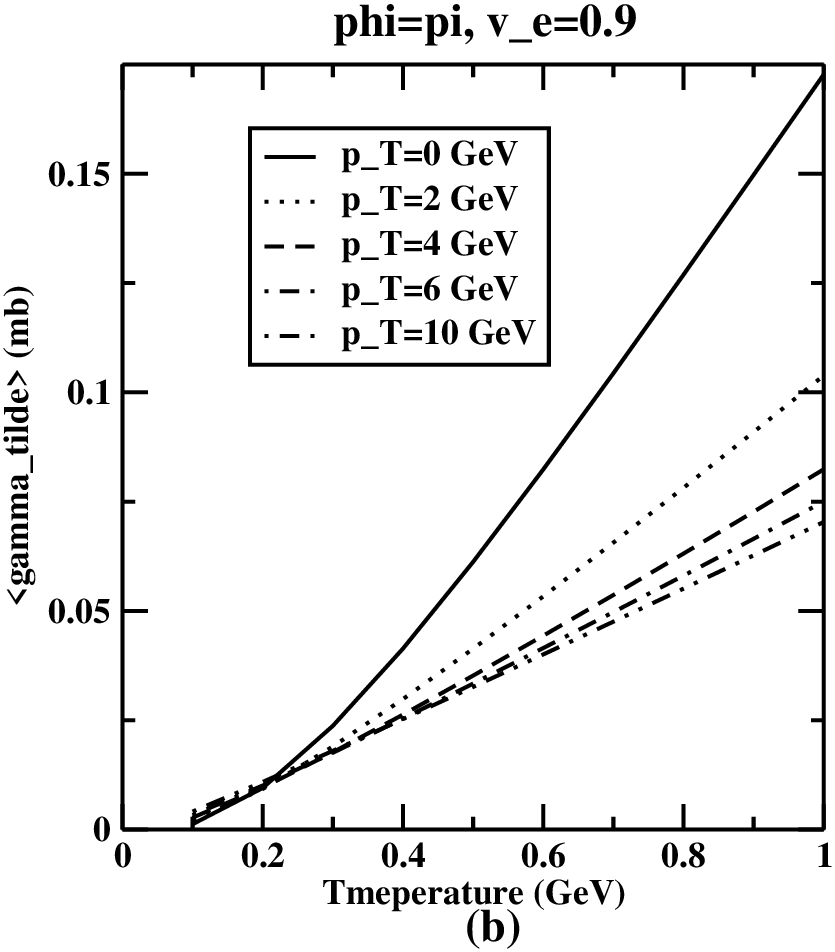,width=7cm}}
\vskip0.5cm
\caption{The variation of the modified rate $\tilg$ as a function of 
temperature at different transverse momenta for the transverse flow
velocity $v=0.9~c$ for (a) $\phii=0$ and (b) $\phii=\pi$, respectively.}
\end{figure}
\begin{figure}[htb]
\mbox{
\hskip1cm\epsfig{file=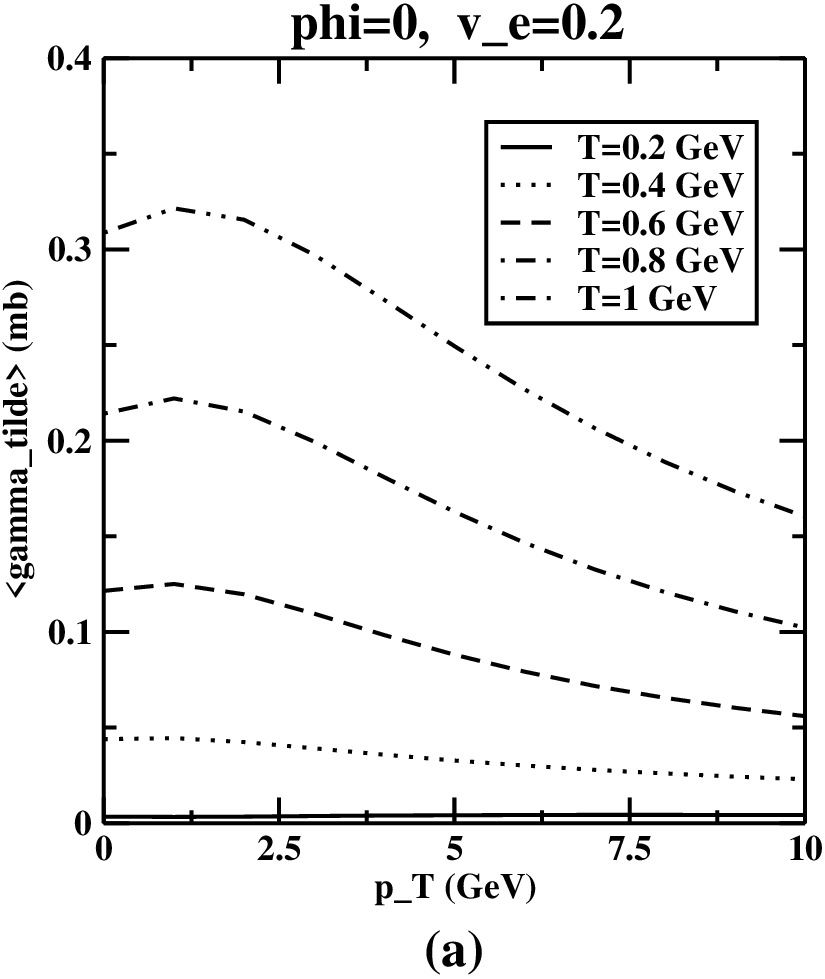,width=7cm}
\hskip1cm
\epsfig{file=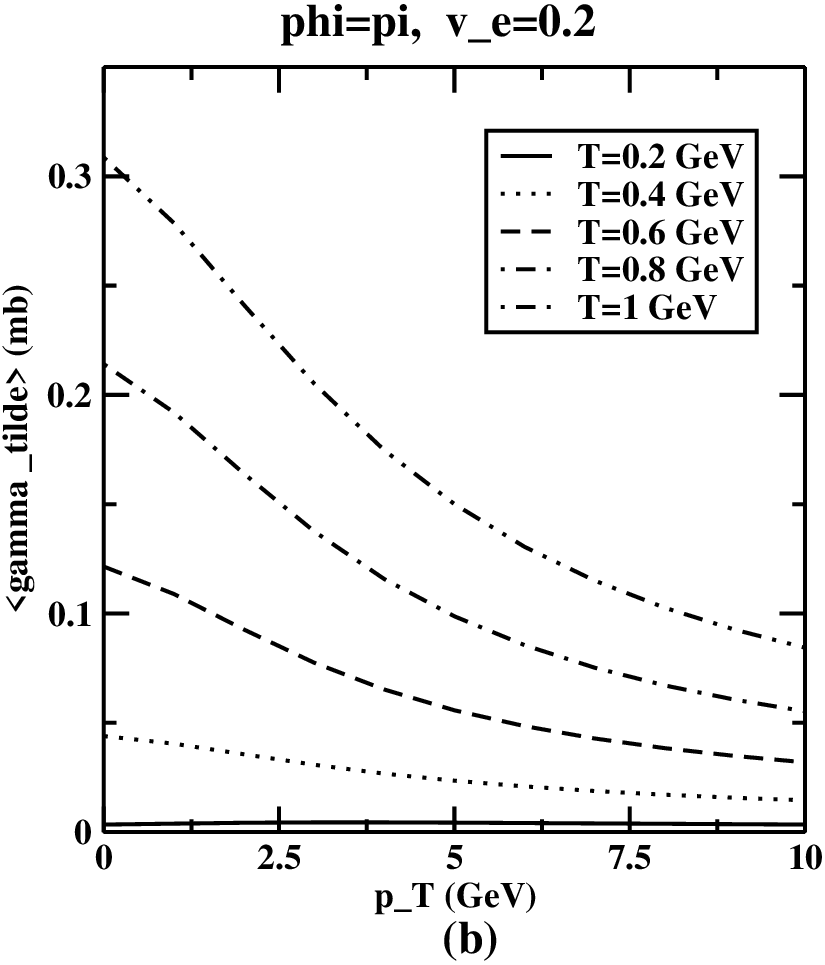,width=7cm}}
\vskip0.5cm
\caption{The variation of the modified rate $\tilg$ as a function of 
transverse momentum for different values of temperatures for a transverse flow
velocity $v=0.2~c$ for (a) $\phii=0$ and (b) $\phii=\pi$, respectively.}
\end{figure}
\begin{figure}[htb]
\mbox{
\hskip1cm\epsfig{file=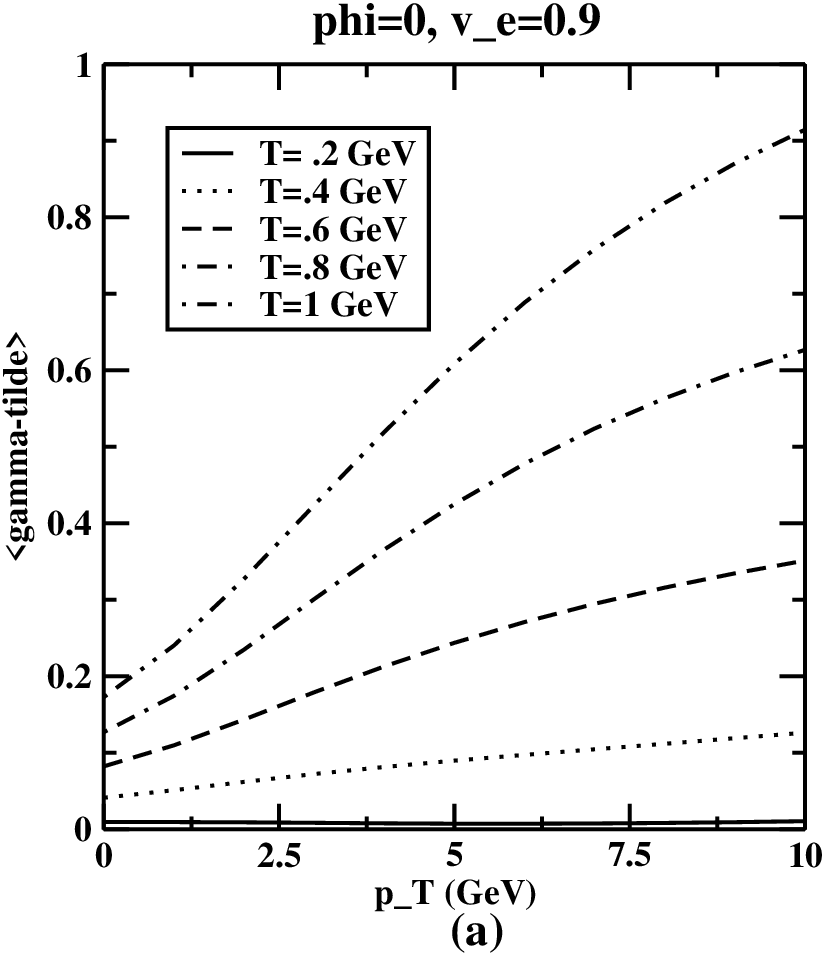,width=7cm}
\hskip1cm
\epsfig{file=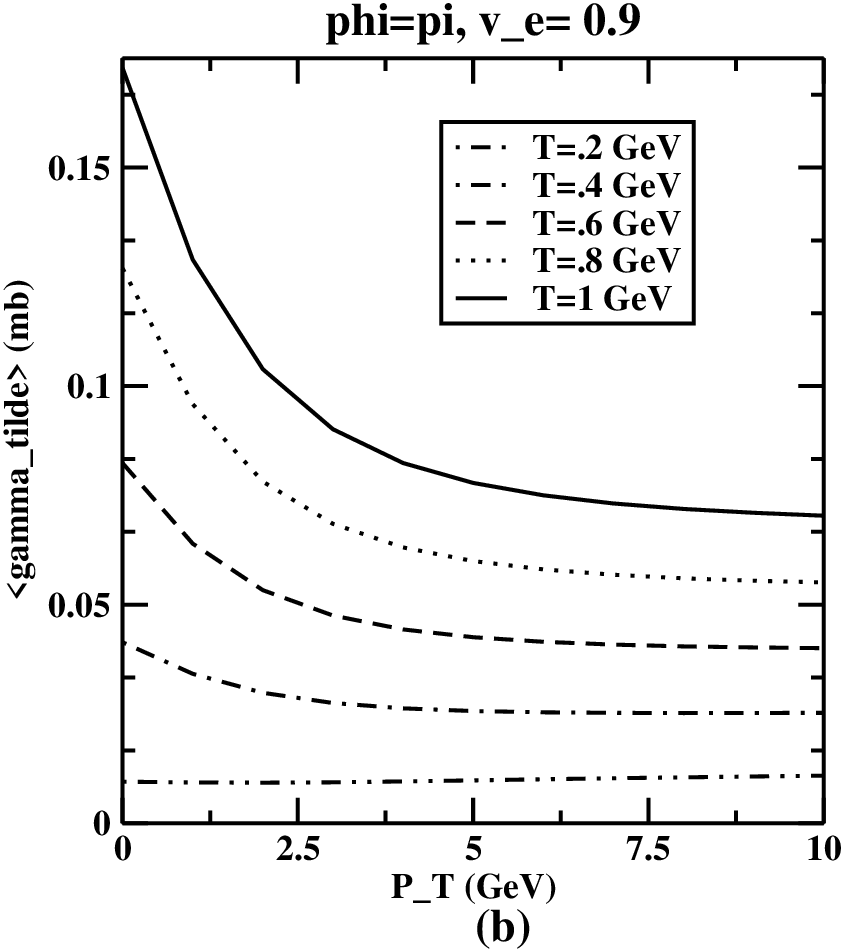,width=7cm}}
\vskip0.5cm
\caption{The variation of the modified rate $\tilg$ as a function of 
transverse momentum at different values of temperatures for a transverse flow
velocity $v=0.9~c$ for (a) $\phii=0$ and (b) $\phii=\pi$, respectively.}
\end{figure}
ii) At fixed value of $(T, \phii, \vex=0.2)$
corresponding to {\it nonrelativistic} flow the variation of $\tilg$
with $\pt$ in Figs.3a,b is more intricate. At $\phii=0$ in Fig.3a
there is a broad {\it enhancement} of $\tilg$ for low $\pt \le 1$ GeV;
this is because, firstly low speeds of the $\psi$ and plasma can compete,
and secondly constructive interference occurs between $I_0$ and $I_1$
in the estimate (2.12) for $\cos \theta_{\psi w}=+1$ ({\em cf.}(3.2)).
On the other hand, at $\phii =\pi$ in Fig.3b our $\tilg$ {\it decreases}
monotonically with $\pt$ throughout; this is due to the fact that,
since $\cos \theta_{\psi w}=-1$ now ({\em cf.}(3.2)), the interference
between $I_0$ and $I_1$ becomes destructive. iii) At fixed values
of $(T, \phii, \vex=0.9)$ corresponding  to {\it ultrarelativistic}
flow similar trends with respect to $\pt$ are again explained in Figs.4a, b
except for the fact that the steady {\it rise} of $\tilg$ with
$\pt$ in Fig.4a is caused mainly by the $\gamma_\psi$ coefficient
present in the estimate (2.12).

\subsection{\bf Curves of survival probability}
For a chosen creation configuration of the $\psi$ meson the function
$W$ was first computed from (2.18) and then $\spt$ was numerically evaluated. 
Figures 5a and 5b show the dependence of $\spt$ on $\pt$ corresponding to the
LHC and RHIC initial conditions, respectively (for two choices of the
transverse expansion speed $\vex$). For the sake of direct comparison, we
also include our earlier results based on no flow ~\cite[eq.(25)]{gluon1} and
longitudinal expansion~\cite[eq.(52)]{gluon2} (starting from two possible 
lengths $L_i$ of the
initial cylinder). We now turn to a physical discussion of these graphs.

\begin{figure}[htb]
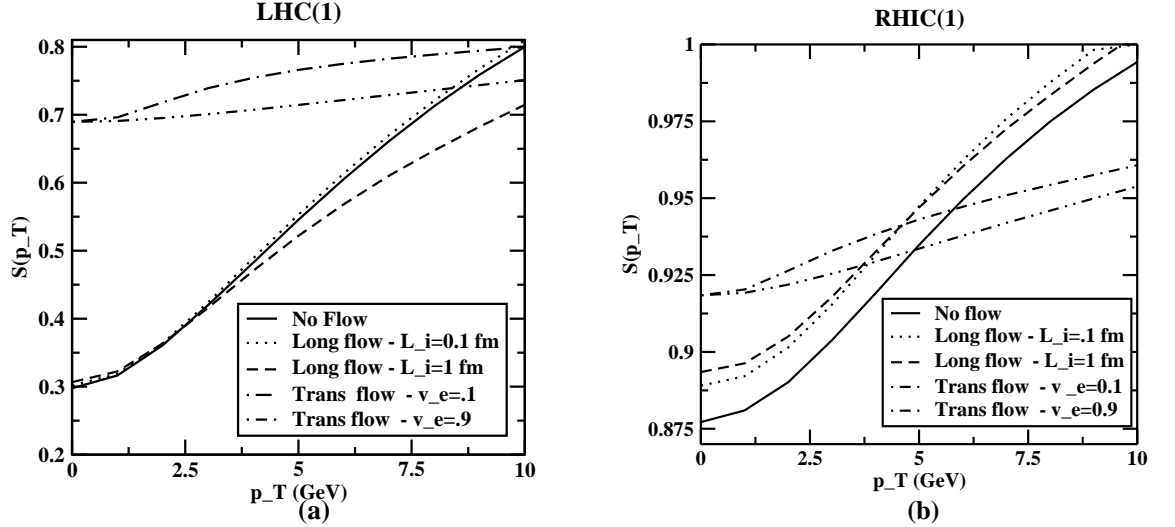

\mbox{
\hskip1cm\epsfig{file=lhc_tr.eps,width=7cm}
\hskip1cm
\epsfig{file=rhic_tr.eps,width=7cm}}
\vskip0.5cm
\caption{The survival probability of $J/\psi$ in an equilibrating
parton plasma at (a) LHC(1)  and (b) RHIC(1) energies with initial conditions 
given in Table 1. The solid curve $\sopt$ is the result 
of~\protect \cite{gluon1}, i.e., in the {\em absence of flow}
while the dotted and dashed curves represent 
the $\sppt$ when the plasma is undergoing longitudinal expansion
with the initial values of the length of the cylinder $L_i=0.1$ fm and
$1$ fm, respectively~\protect\cite{gluon2}. The dot-dashed and and double 
dot-dashed curves depict the $\stpt$ when the system is undergoing
transverse expansion with the expansion speed $\vex=0.1$ and $0.9$,
respectively. }
\end{figure}

\nd {\bf Interpretation}:
In every scenario of gluonic  dissociation the function $W=
\int_{t_I}^{t_{II}} dt~\tilg $ depends on $\pt$ via the integrand
$\tilg$ as well as the limits $(t_I, t_{II})$. Three interesting
cases may now be distinguished:

\nd {\it No flow case}~\cite{gluon1}: Here cooling of the plasma is simulated 
through the master rate equations but the existence of the explicit flow
profile is ignored. Then $\tilg$ decreases monotonically with $\pt$
because of a destructive interference between the $I_0$ and $I_1$ terms.
Also, the time-span $\tsp$ is shortened  as the speed of the
$\psi$ meson increases. Consequently, the survival probability called $\sopt$
grows steadily with $\pt$ as shown by the solid lines in Figs.5a,b.

\nd {\it Longitudinal expansion case}: Here an extra parameter appears namely 
the
length $L_i$ of the initial cylinder. For nonrelativistic flow emanating
from short length $L_i=0.1$ fm the $\tilg$ values are somewhat reduced 
compared to the no flow case (due to $I_0$, $I_1$ destructive interference)
though the time span $\tsp$ remains unaltered, so that the
survival probability called $\sppt$ is pushed slightly upwards in Figs.5a,b.
But for relativistic flow emanating from longer length $L_i=1$ fm
the shifts of the $\sppt$ curve occurs in mutually opposite directions
at LHC and RHIC (due to the different initial temperatures generated
therein). 

\nd {\it Transverse expansion case}: Here the extra parameter involved is the
transverse expansion speed $\vex$ which together with $\phii$ and $T$
control the $\pt$-dependence of the function $W$. For $\phii=0$ the
$\tilg$ values in Figs.3a, 4a exhibit enhancement/rising trend on the lower
$\pt$ side; such $\psi$ mesons contribute sizably to $W$ but little
to $e^{-W}$. On the othe hand, all curves of $\tilg$ in Figs.3 - 4
flatten-off to nearly constant values on the higher $\pt$ side; such
$\psi$ mesons contribute substantially to $e^{-W}$ especially for
low temperatures. Therefore, the transverse survival probability $\stpt$
becomes nearly $\pt$-independent (or very slowly varying) in Figs.5a, 5b
in sharp contrast to the longitudinal case. For explaining the magnitude of the
ratio $\stpt/\sppt$ we consider the temporal scenario dealing with the
limits of the integration.

{\it Temporal scenario}: It is known that transverse expansion of a quark-gluon
plasma produces cooling at a faster rate compared to longitudinal expansion
so that the inequality $t^\perp_{\mbox{life}} <t^\parallel_{\mbox{life}}$
holds on the corresponding lifetimes. At LHC the transverse cooling is so
fast that, for most $\psi$ mesons of kinematic interest we have $t_{II}
=\tlf^\perp$ in the definition (2.15). The time-span $\tsp$ is, therefore, much
smaller compared to the longitudinal case implying thet $\stpt > \sppt$ in
Fig.5a. Clearly this property at LHC is devoid of any rich structure.

However, at RHIC let us divide the $\psi$ meson kinematic region into two
parts. For slower mesons having $\pt<5$ GeV the catch-up time 
$t_I+\Delta^\ast$ in (2.15) exceeds the lifetime so that $t_{II}=\tlf^\perp$
again, i.e., $\stpt > \sppt$ in Fig.5b for $\pt < 5$ GeV. Next, for faster 
mesons
having $\pt>5$ GeV the reverse inequalities hold making $\stpt < \sppt$
in Fig.5b. Clearly, the rich structure in $\stpt$ at RHIC arises from a
mutual competition between the catch-up time and lifetime.

\section{\bf Conclusions}
a) In this work we have applied our general formulation~\cite{gluon2} of
hydrodynamic expansion to study the effect of explicit transverse flow profile
on the gluonic break up of $\jsi$'s created in an equilibrating QGP. The 
formalism in Sec.2 and numerical results of Sec.3 are new and original.

b) Equation (2.8) shows that, at specified fugacity $\lamg$, the effect of 
transverse flow is to increase the gluon number density $\ng$. This was also
the case with longitudinal flow.

c) Our expressions (2.10, 2.12) of the mean dissociation rate $\tilg$ 
involves hyperbolic functions as well as partial wave interference
mechanism (controlled by the anisotropic $\cos \theta_{\psi w}$ factor).
In addition, knowledge about a nontrivial kinematic function $F$ 
({\em cf.}(2.16)) is needed for interpreting the variation of $\tilg$ with $T$, $\pt$,
$\phii$, $\vex$ in Figs.1 - 4. In contrast, for longitudinal flow the
treatment of $\tilg$ was easier because $F=0$ there.

d) There are several features of contrast between the transverse and 
longitudinal survival probabilities denoted by $\stpt$ and $\sppt$,
respectively. Due to the geometry of production configuration our
$\stpt$ contains a double integral (2.18) whereas $\sppt$ contains a triple
integral. Next, due to the flattening-off trend of $\tilg$ with increasing
$\pt$ our $\stpt$ becomes roughly $\pt$-independent (or slowly varying) in 
Figs.5a, 5b
whereas $\sppt$ rises rapidly. Finally, the quick cooling rate at LHC
makes $\stpt > \sppt$ at all $\pt$ of interest in Fig.5a whereas a 
competition between the catch-up time and lifetime generates richer
structure at RHIC in Fig.5b.

e) We conclude with the observation that the field of $\jsi$ suppression due 
to gluonic break up continues to be a research area of great challenge. In
a future communication we plan to study the effect of asymmetric flow profile
arising from noncentral collision of heavy ions at finite
impact parameter $\vec{b}$.

\section*{ACKNOWLEDGEMENTS}  
VJM thanks the UGC, Government of India, New Delhi for financial support.
We are also thankful to Dr. Dinesh Kumar Srivastava for discussing in the
early stages of this work.

\end{document}